\documentclass[prd,twocolumn,superscriptaddress,nofootinbib,showpacs]{revtex4}

\usepackage{graphicx,amsmath,amsfonts,amssymb,revsymb,dcolumn,epsfig,bm}

\begin{document}

\title{G\"{o}del-type universes in Palatini $\mathbf{f(R)}$ gravity}

\author{J. Santos}
\affiliation{Universidade Federal do Rio G. do Norte,
Departamento de F\'{\i}sica,  \\
59072-970 Natal -- RN, Brazil}

\author{M.J. Rebou\c{c}as}\email{reboucas@cbpf.br}
\affiliation{Centro Brasileiro de Pesquisas F\'{\i}sicas,
Rua Dr.\ Xavier Sigaud 150,  \\
22290-180 Rio de Janeiro -- RJ, Brazil}

\author{T.B.R.F. Oliveira}
\affiliation{Universidade Federal do Rio G. do Norte,
Departamento de F\'{\i}sica,  \\
59072-970 Natal -- RN, Brazil}

\date{\today}

\begin{abstract}
We examine the question as to whether the Palatini f(R) gravity
theories permit space-times in which the causality is violated. We show that
every perfect-fluid G\"{o}del-type solution of Palatini f(R) gravity with
density $\rho$ and pressure $p$ that satisfy the weak energy condition $\rho+p
\geq 0$ is necessarily isometric to the G\"odel geometry, demonstrating
therefore that these theories present causal anomalies in the form of closed
time-like curves. This result extends a theorem on G\"{o}del-type models to the
framework of Palatini f(R) gravity theory. We concretely examine the
G\"odel-type perfect-fluid solutions in specific $f(R) = R - \alpha/R^{n}$
Palatini gravity theory, where the free parameters $\alpha$ and $n$ have been
recently constrained by observational data. We show that for
positive matter density and for $\alpha$ and $n$ within the interval permitted
by the observations, this theory does not admit the G\"odel geometry as a
perfect-fluid solution of its field equations. In this sense, this theory
remedies the causal pathology in the form of closed time-like curves which is
allowed in general relativity. We derive an expression for a critical radius
$r_c$ (beyond which the causality is violated) for an arbitrary Palatini f(R)
theory. The expression makes apparent that the violation of causality depends
on the form of f(R) and on the matter content components. We also examine the
violation of causality of G\"odel-type by considering a single scalar field as
the matter content. For this source we show that Palatini f(R) gravity gives
rise to a unique G\"odel-type solution with no violation of causality.
\end{abstract}


\maketitle

\section{Introduction}

The $f(R)$ gravity theory provides an alternative way to
explain the current  cosmic acceleration with no need of invoking
either a dark energy component or the existence of an extra spatial
dimension.
The freedom in the choice of different functional forms of $f(R)$,
however, gives rise to the problem of how to constrain on theoretical
and/or observational grounds, the many possible $f(R)$ gravity theories.
A great deal of effort has gone into the study of some
features of these theories~\cite{theory} (see also Refs.~\cite{Francaviglia}
for recent reviews). This includes solar system tests \cite{Chiba2003}, Newtonian
limit~\cite{Sotiriou2006a}, gravitational stability \cite{Dolgov2003}
and singularities~\cite{Frolov2008}.
General principles such as the so-called energy conditions have also been
used to place constraints on $f(R)$ theory~\cite{energy_conditions}.
Recently, observational constraints from several cosmological data sets
have also been employed to test the viability of some $f(R)$ theories~\cite{Amarzguioui,Tavakol,Carvalho,Santos,Li,Yang,Borowiec}.

In dealing with $f(R)$ gravity theories two different variation approaches
may be followed, namely  the metric  and the Palatini formalisms.
In the metric approach the connection is  assumed to be the Levi-Civita
connection,  and variation of the action is taken with respect to the metric,
whereas in the Palatini approach the metric and the affine connections are
treated as independent fields and the variation of the action is taken with
respect to both metric and connections.
Although these approaches lead to the same set of field equations in the
context of general relativity (GR), for a general $f(R)$ with non-linear
term in the Einstein-Hilbert action they give rise to different
field equations. In this paper we shall focus on $f(R)$ gravity in the
Palatini formalism.

In both versions of the $f(R)$ gravity theories the causal structure of
four-dimensional space-time has locally the same qualitative nature as
that of the flat space-time of special relativity --- causality holds locally.
The nonlocal question, however, is left open, and violation of
causality can occur. However,  if gravity is governed by a $f(R)$
theory instead of GR, various issues of both observational and theoretical
nature ought to be reexamined in the $f(R)$ gravity framework, including
the question as to whether these theories  permit
space-time solutions of their field equations in which the causality
is violated.

In general relativity there are solutions to the field equations that
have causal anomalies in the form of closed time-like curves.
The renowned G\"odel model~\cite{Godel49} is the best
known example of such a  solution, which makes apparent
that the GR does not exclude the existence of solutions with
closed timelike world lines, despite its Lorentzian character
that leads to the local validity of the causality principle.
The G\"odel model is a solution of Einstein's equations
with cosmological constant $\Lambda$ for dust of density
$\rho$, but it can also be interpreted as perfect-fluid
solution (with pressure $p=\rho\,$) without cosmological
constant.
In this regard, we recall that it was shown by Bampi and
Zordan~\cite{BampiZordan78} that every G\"odel-type solution
of Einstein's equations with a perfect-fluid energy-momentum
tensor is necessarily isometric to the G\"odel spacetime.

Owing to its unexpected properties, G\"odel's model has a
well-recognized importance and has motivated a 
number of investigations on rotating G\"odel-type models
as well as on causal anomalies not only in the context of general
relativity (see, e.g. Refs.~\cite{GT_in_GR})
but also in the framework of other theories of gravitation
(see, for example, Refs.~\cite{GT_Other_Th}).

In a recent paper, we have examined G\"odel-type models and the violation
of causality problem  for $f(R)$ gravity in the \emph{metric} variational
approach~\cite{RS-2009}, generalizing therefore the results of
Refs.~\cite{CliftonBarrow2005} and~\cite{Reb_Tiomno}.
In this article, to proceed further with the investigation of G\"odel-type
universes along with the question of breakdown of causality in $f(R)$ gravity,
we extend the results of Refs.~\cite{RS-2009,CliftonBarrow2005,Reb_Tiomno}
by examining the question as to whether the $f(R)$ gravity theories in the
\emph{Palatini} formalism admit G\"odel-type space-times solutions
in which violation of causality can occur for a physically
well-motivated matter source. In this way, we extend the results of
Ref.~\cite{RS-2009,Reb_Tiomno} in four different regards.
First, we demonstrate that every perfect-fluid G\"{o}del-type solution of
any  Palatini $f(R)$ gravity with density
$\rho$ and pressure $p$ and satisfying the weak energy condition
$\rho+p \geq 0$ or equivalently $df/dR > 0$%
\footnote{In the metric formalism,  this condition is necessary to ensure that
the effective Newton constant $G_{\text{ef\/f}} = G / f_R$ does not change its sign.
At a quantum level, it prevents the graviton from becoming ghostlike.}
is necessarily isometric to the G\"odel geometry.
This extends to the context of Palatini $f(R)$ the so-called
Bampi-Zordan theorem~\cite{BampiZordan78} which was
established in the context of Einstein's theory, and has been extended
recently to the framework of  $f(R)$ in the \emph{metric} formalism~\cite{RS-2009}.
Second, we examine the dependence of the critical radius $r_c$ (beyond which
the causality is violated) with both the  Palatini $f(R)$ gravity and
the fluid components ($p$ and $\rho$) and derive an expression for
$r_c$ that holds for any Palatini $f(R)$ gravity theory.
Third, we concretely illustrate our general results for perfect-fluid
G\"odel-type solutions in Palatini $f(R)$ gravity by taking the
specific $f(R) = R - \alpha/R^{n}$ theory, where the free parameters
$\alpha$ and $n$ have been recently constrained by a diverse set of
observational data.
We show that for positive matter density and for $\alpha$ and $n$
within the interval permitted by the observational data, this theory
does not admit G\"odel geometry as a perfect-fluid solution of its
field equations.
In this sense, this theory remedies the causal pathology in the form of
closed time-like curves which is allowed in general relativity.
Fourth,  we  examine the violation of causality of G\"odel type
by considering a scalar field as a matter source. For this source
we show that Palatini $f(R)$ gravity gives rise to a unique
G\"odel-type solution with no violation of causality.

\section{$\mathbf{f(R)}$ gravity in the Palatini approach}

The causality problem in $f(R)$ gravity theories can be seen as
having three interrelated physically determinant ingredients,
namely the gravity theory, the space-time geometry and the
matter source.
Regarding the first ingredient we recall that the action that
defines a $f(R)$ gravity is given by
\begin{equation}
\label{actionJF}
S = \int d^4x\sqrt{-g}\left[ \frac{f(R)}{2\kappa^2} + \mathcal{L}_m \right]\,,
\end{equation}
where $g$ is the determinant of the metric tensor $g_{\mu \nu}\,$, $f(R)$ is
a function of the Ricci scalar $R$, $\kappa^2=8\pi G$, and $\mathcal{L}_m$ is
the Lagrangian density for the matter fields.
Treating the metric and the connection as independent fields, the variation of
this action with respect to the metric gives the field equations
\begin{equation}
\label{field_eq}
f_R R_{(\mu\nu)} - \frac{f}{2}g_{\mu\nu}  = \kappa^2T_{\mu\nu}\,,
\end{equation}
where $f_R=df/dR$, $T_{\mu\nu} = -(2\,/\!\sqrt{-g})\,\,\delta (\sqrt{-g}\mathcal{L}_m )
/ \, \delta g^{\mu\nu}$ is the matter energy-momentum tensor, and
$R_{\mu\nu}$ is given in the usual way in terms of
the connection $\Gamma_{\mu\nu}^{\rho}$ and its derivatives.

The variation of the action (\ref{actionJF}) with respect to the
connections field yields
\begin{equation}  \label{connections_eq}
\widetilde{\nabla}_\beta\left( f_R \, \sqrt{-g}\,g^{\mu\nu}\right) = 0 \,,
\end{equation}
where  $\widetilde{\nabla}_\beta$ denotes the covariant derivative
associated with the $\Gamma_{\mu\nu}^{\rho}$.
If one defines a metric $h_{\mu\nu} = f_R (R)\, g_{\mu\nu}$ it can be easily
shown that Eq.~(\ref{connections_eq}) determines a Levi-Civita connection of
$h_{\mu\nu}$, which in turn can be rewritten in terms of $g_{\mu\nu}$ and
its  Levi-Civita connection in the form
\begin{equation}  \label{Gamma}
\Gamma_{\mu\nu}^{\rho} = \left\{^{\rho}_{\mu\nu}\right\}
+ \frac{1}{2f_R}\left( \delta^{\rho}_{\mu}\partial_{\nu}
+ \delta^{\rho}_{\nu}\partial_{\mu}
- g_{\mu\nu}g^{\rho\sigma}\partial_{\sigma} \right)f_R\,.
\end{equation}
An important constraint, often used to simplify the field equations, comes
from the trace of equation (\ref{field_eq}), which is  given by
\begin{equation}  \label{trace_eq}
f_R \, R(\Gamma) - 2f = \kappa^2\,T\,,
\end{equation}
where $T=g^{\mu\nu}T_{\mu\nu}$ is the trace of the energy-momentum tensor
and $R(\Gamma)=g^{\mu\nu}R_{\mu\nu}$ is calculated with the
connection $\Gamma_{\mu\nu}^{\rho}$ given by Eq.~(\ref{Gamma}).

In practice, it turns out to be useful to express the field
equations~(\ref{field_eq}) in terms of the metric $g_{\mu\nu}$,
its derivatives, and the matter fields. To this end,
one uses equations~(\ref{Gamma}) and~(\ref{trace_eq}) to
eliminate the connection $\Gamma_{\mu\nu}^{\rho}$ from the field
equations~(\ref{field_eq}). After some manipulations one obtains
\begin{eqnarray} \label{einstein-like_eq}
f_R\,G_{\mu\nu} & = & \kappa^2\,T_{\mu\nu} - \frac{1}{2}(\kappa^2\,T
+ f)\,g_{\mu\nu} + H_{\mu\nu}\,f_R \nonumber \\
 &  & - \frac{3}{2f_R}\left[ \nabla_{\mu}f_R\nabla_{\nu}f_R
 - \frac{1}{2}g_{\mu\nu}(\nabla f_R)^2 \right],
\end{eqnarray}
where $\nabla_\mu$ denotes the covariant derivative associated with
the  Levi-Civita connection of the metric $g_{\mu\nu}\,$,
$\Box = g^{\alpha \beta}\,\nabla_{\alpha}\nabla_{\beta}\,$, %
$H_{\mu\nu}\equiv \nabla_{\mu}\nabla_{\nu} - g_{\mu\nu}\,\Box$, and
$G_{\mu\nu}=R_{\mu\nu} - R/2\,g_{\mu\nu}$ is the Einstein tensor, which
is also calculated with the metric Levi-Civita connection.

Having given an account of the first basic ingredient of the causality
problem, i.e. $f(R)$ gravity in the Palatini approach, in the next section
we shall examine the second relevant component of this problem, which is
the G\"odel-type geometries, and discuss how the violation of causality can
occur in G\"odel-type spacetimes.

\section{G\"{o}del-type Geometries} \label{Goedel-geo}

The G\"odel-type class of geometries that we focus our attention on
in this article is given, in cylindrical coordinates $(r, \phi, z)$,
by~\cite{Reb_Tiomno}
\begin{equation}  \label{G-type_metric}
ds^2 = [dt + H(r)d\phi]^2 - D^2(r)d\phi^2 - dr^2 - dz^2\,,
\end{equation}
where
\begin{eqnarray}
H(r) & = & \frac{4\omega}{m^2}\,\sinh^2(\frac{mr}{2})\,, \label{godel_funH} \\
D(r) & = & \frac{1}{m}\,\sinh(mr)\,, \label{godel_funD}
\end{eqnarray}
where $\omega$ and $m$ are constant parameters such that $\omega^2 > 0$ and
$-\infty\leq m^2\leq +\infty$.
Clearly, for $m^2 = - \mu^2 <0$ the metric
functions $H(r)$ and $D(r)$ become circular functions
$H(r)=(4\omega/\mu^2)\sin^2(\mu r/2)$ and
$D(r)=\mu^{-1} \sin(\mu r)$, while in the limiting case $m=0$
they become $H= \omega\, r^2$ and $D = r$.

All G\"odel-type geometries are characterized by the
two parameters $m$ and $\omega$. In this way,  identical pairs $(m^2, \omega^2)$
specify isometric spacetimes~\cite{Reb_Tiomno,TeiRebAman,RebAman}. G\"odel solution
is a particular case of the $0 < m^2 < +\infty$ class of spacetimes in which
$m^2= 2 \omega^2$.

In order to examine the causality features of G\"odel-type
we first note that the G\"odel-type line element~(\ref{G-type_metric})
can be rewritten as
\begin{equation} \label{G-type_metric2}
ds^2=dt^2 +2\,H(r)\, dt\,d\phi -dr^2 -G(r)\,d\phi^2 -dz^2 \,,
\end{equation}
where $G(r)= D^2 - H^2$. In this form it is clear that the circles
defined by $t, z, r = \text{const}$, are closed timelike curves
depending  on the sign of $G(r)$.
Thus, for $G(r) < 0$ for a certain range of $r$ ($r_1 < r < r_2$, say)
the so-called \emph{G\"odel circles} defined by  $t, z, r = \text{const}$
are closed time-like curves.
By using this inequality along with the equations (\ref{godel_funH})
and~(\ref{godel_funD}) it is  easy to show that the causality
features of the  G\"odel-type space-times depend upon
the two independent parameters $m$ and $\omega$
as it follows~\cite{Reb_Tiomno}.
For $m=0$ there is a critical radius, $r_c = 1/\omega$,
such that for all $r>r_c$ there are noncausal G\"odel circles
defined by $t, z, r = \text{const}$.
For $m^2= -\mu^2 <0$ the functions $H$ and $D$ are trigonometric functions
and there is an infinite sequence of alternating causal and noncausal
$t, z, r = \text{const}$ regions without and with G\"odel circles.
For $0 < m^2 < 4\omega^2$ noncausal
G\"odel circles occur for $r>r_c$ such that
\begin{equation} \label{r-critical}
\sinh^2 \frac{mr_c}{2}= \left[ \frac{4\omega^2}{m^2} - 1 \right]^{-1}.
\end{equation}
When  $m^2 = 4 \omega^2$ the critical radius $r_c \rightarrow \infty$.
Thus, for $m^2 \geq 4 \omega^2$ there are no G\"odel circles, and hence
the breakdown of causality of G\"odel-type is avoided.

To close this section, we note that the presence of a single
closed timelike curve as, for example,  a G\"odel's circle,
is an unequivocal manifestation of violation of causality.
However, a space-time may admit noncausal closed curves other than
G\"odel's circles. In this paper, by non-causal and
causal solutions we mean, respectively, solutions with and
without violation of causality of G\"odel-type, i.e., with
and without G\"odel's circles. Clearly this type of violation
of causality is not of trivial topological nature, which are
obtained by topological identification~\cite{Topology}.

\section{G\"{o}del-type solutions in Palatini  $\mathbf{f(R)}$  gravity}

The third important ingredient in the above mentioned causality problem
is the matter source, which we shall discuss in this section. To this
end, we first show  how the field equations~(\ref{einstein-like_eq})
can be greatly simplified for G\"odel-type geometries, and then
we discuss the role played by two matter sources in the breakdown of
causality of G\"odel-type.

\subsection{Field equations}

From Eqs.~(\ref{G-type_metric}), (\ref{godel_funH}) and (\ref{godel_funD})
it is straightforward to show that the Ricci scalar for the G\"odel-type
metrics takes a constant value $R = 2 (m^2 - \omega^2)$. Hence, the last three
terms of the field equations~(\ref{einstein-like_eq}) vanish. A further
simplification  comes about if instead of using coordinates
basis one uses the following locally Lorentzian basis:
\begin{eqnarray}
\theta^0 &=& dt + H(r)d\phi\,, \quad  \theta^1 = dr\,, \label{one_forms1} \\
\theta^2 &=& D(r)d\phi\,,        \,\quad \qquad \theta^3 = dz \label{one_forms2} \,,
\end{eqnarray}
relative to which the G\"odel-type line element~(\ref{G-type_metric}) clearly takes
the form
\begin{equation}  \label{G-type_metric3}
ds^2 = \eta_{AB}\,\theta^A\theta^B =
(\theta^0)^2 - (\theta^1)^2 - (\theta^2)^2 - (\theta^3)^2\,.
\end{equation}
In this basis the field equations~(\ref{einstein-like_eq}) reduce
to
\begin{equation}  \label{G_AB}
f_R\,G_{AB} = \kappa^2\,T_{AB} - \frac{1}{2}\left( \kappa^2T + f \right)\eta_{AB}\,,
\end{equation}
where the nonvanishing components of the Einstein tensor $G_{AB}$ take
the quite simple form
\begin{equation} \label{GAB_components}
G_{00} =  3 \omega^2 - m^2, \;\,
G_{11} = G_{22}  =  \omega^2, \;\,
G_{33}  =  m^2 - \omega^2\,.
\end{equation}

In the next subsections we examine whether these theories permit causal
and non-causal solutions for two physically well-motivated matter
sources, namely a perfect fluid and a single scalar field.

\subsection{Perfect fluid solutions}

We first consider a perfect-fluid of density $\rho$ and pressure $p$,
whose energy-momentum tensor in the basis~(\ref{one_forms1})
and~(\ref{one_forms2}) is clearly given by
\begin{equation}  \label{perfect_fluid}
T^{(M)}_{AB} = (\rho + p)\,u_{A}u_{B} - p\,\eta_{AB}\,.
\end{equation}
Taking into account Eq.~(\ref{GAB_components}) and  Eq.~(\ref{G_AB}), for
this matter source the field equations reduce to
\begin{eqnarray}
2\omega^2f_R - f &=& \kappa^2\,(\rho - p)\,, \label{1-eq} \\
2(m^2 - \omega^2)f_R - f &=& \kappa^2\,(\rho - p)\,,\label{2-eq} \\
2(3\omega^2 - m^2)f_R + f &=& \kappa^2\,(\rho + 3p)\,.\label{3-eq}
\end{eqnarray}
From Eqs.~(\ref{1-eq}) and (\ref{2-eq}) we obtain
$f_R \,(m^2 - 2\omega^2) = 0$.
Thus, for $f(R)$ theories that satisfy the week energy condition $f_R> 0$
(see next paragraph for details) 
we have $m^2 = 2 \omega^2$, which according to Sec.~\ref{Goedel-geo}
defines the G\"odel metric.
Thus, a general class of perfect fluid G\"odel-type solutions
of Palatini $f(R)$  gravity is given by
\begin{eqnarray}
m^2 & = & 2\omega^2\,, \label{m-eq} \\
\kappa^2 \rho &=& m^2 f_R - \frac{f}{2}  \,, \label{rho-eq} \\
\kappa^2 p &=& \frac{f}{2} \,,\label{p-eq}
\end{eqnarray}
where $f$ and $f_R$ are evaluated at $R=2(m^2-\omega^2)=m^2$.

Now, recalling the week energy condition (WEC)~\cite{EC} takes the form%
\footnote{For studies of the interrelations between the energy
conditions (on scales relevant for cosmology) and observational
data see Ref.~\cite{EC-observation}.}
$\rho \geq  0$ and $\rho+p \geq 0$,  it is clear from~(\ref{rho-eq})
and~(\ref{p-eq}) that $m^2 f_R = \kappa^2 (\rho+p)$,  and thus
the second  WEC inequality is identically satisfied for any Palatini
$f(R)$ gravity theories with $f_R > 0\,$.
In this way, equations~(\ref{m-eq}),  (\ref{rho-eq}) and~(\ref{p-eq})
show that  the G\"odel geometry arises as perfect fluid solution
of any Palatini $f(R)$  gravity in which  $\rho+p > 0$.
This result can be looked upon as an extension of Bampi and
Zordan~\cite{BampiZordan78} theorem%
\footnote{Bampi and Zordan theorem was obtained originally
in the framework of general relativity and extended to the
context of $f(R)$ gravity in the metric formalism in
Ref.~\cite{RS-2009}.}
to the context of Palatini $f(R)$ gravity in the sense that
for arbitrary $\rho$ and $p$ (with $\rho+p >0$) all
perfect-fluid  G\"odel-type solution of every Palatini $f(R)$  gravity,
which satisfies the condition $f_R>0$, are necessarily isometric to the
G\"odel geometry.

Regarding the causality properties of this general family of perfect-fluid
G\"odel-type solutions, we first note that since they  are isometric to
G\"odel geometry they admit noncausal G\"odel circles of radius greater
than the critical radius $r_c$ given by Eq.~(\ref{r-critical}). But,
taking into account Eqs.~(\ref{rho-eq}) and~(\ref{p-eq}) we have
now that
\begin{equation}  \label{critical_radius}
r_c = 2 \,  \sqrt{ \frac{f_R^{}}{\kappa^2(\rho + p) } } \; \,\sinh^{-1} (1)\,\,,
\end{equation}
making apparent that the critical radius, beyond which there exist
noncausal G\"odel circles, depends on the two physically relevant ingredients,
namely the gravity theory and the matter source components, as one
would expect from the outset.

Before proceeding, some words of clarification are in order concerning
the first inequality of the WEC, which ensures the  positivity of
the matter density $\rho$.
In general relativity [$f(R)=R$] \ Eqs.~(\ref{p-eq}) and~(\ref{rho-eq})
clearly yield $\kappa \rho = \kappa^2 p = m^2/2$, making clear that
both the matter density and the pressure are positive. However, for
a general $f(R)$ (with non-linear terms in $R$) these
equations do not necessarily lead to $\rho> 0$
for all values  $m^2 = 2\omega^2$. In this way, the above general
result concerning perfect-fluid G\"odel-type solutions may not
hold for some $f(R)$ gravity if one further demands the first WEC
inequality ($\rho>0$), which from Eq.~(\ref{rho-eq}) leads to
\begin{equation} \label{posrho}
m^2 f_R - \frac{f}{2} \geq 0 \,,
\end{equation}
where $f_R \neq 0$ and both $f$ and $f_R$ are evaluated at $R=m^2$.

As a concrete example, we consider the extensively discussed $f(R)$ theory
given by
\begin{equation}  \label{first-f}
f(R) = R - \frac{\alpha}{R^n}\,,
\end{equation}
where $\alpha$ and $n$ are free parameters to be determined by local
gravity constraints and cosmological observations.
In the metric approach the gravity theories of the form~(\ref{first-f}) 
are know to be plagued with problems~\cite{Chiba2003,Dolgov2003,Sotiriou2006a,Amendola}.
In the Palatini approach, however,  combination of a dynamical
autonomous systems analysis (study of the fixed points and
stabilities against perturbations) yields $n>-1$ for $\alpha >0$,
which can be shown to permit cosmological models with radiation-dominated,
matter-dominated and de Sitter phases~\cite{Tavakol}.
Furthermore, recent constraints from a combination of type-Ia supernova (SNe Ia),
baryon acoustic oscillation peak (BAO) and cosmic microwave background
radiation (CMB) shift give $n\in[-0.3,0.3]$ and $\alpha\in[1.3,7.1]$
at $99.7\,\%$ confidence level~\cite{Amarzguioui,Tavakol,Santos,Carvalho,Li}.

For gravity theory of the form~(\ref{first-f}) the positivity of
the energy density~(\ref{posrho}) gives
\begin{equation}  \label{m-cond1}
m^{2n+2} + (2n + 1)\alpha \geq 0\,.
\end{equation}
Now taking into account  the above  dynamical systems constraint on $n$ and
$\alpha$, one has that there are real values for $m$ such that~(\ref{m-cond1})
holds only for $n$ in the interval $n \in (-1,-0.5)$, whose intersection with the
above interval permitted by observations is empty. This makes clear that the
Palatini $f(R)$ gravity~(\ref{first-f}) does not admit G\"odel geometry as
a perfect-fluid solution with $\rho>0$, and for the values of $n$ and
$\alpha$ allowed by dynamical systems along with the above combination
of observational data. In this sense, this theory remedies the causal
pathology in the form of closed time-like curves which is allowed in
general relativity.

\subsection{Single scalar field solutions}

Since any perfect-fluid G\"odel-type solution of Palatini $f(R)$
gravity that is subject to the WEC condition $\rho+p> 0$ is noncausal,
the question as to whether other matter sources could generate
G\"odel-type causal solutions naturally arises.
In this section we shall examine this question by considering
another different matter source, namely a single scalar field
$\Phi(z)$, whose energy momentum tensor is given by
\begin{equation}
T^{(S)}_{AB}= \Phi^{}_{|A}\,\Phi^{}_{|B} - \frac{1}{2}\,\eta^{}_{AB}\,
\Phi^{}_{|M} \,\Phi^{}_{|N}\, \eta^{MN} ,
\end{equation}
where a vertical bar denotes components of covariant derivatives relative to
the local basis $\theta^A = e^{(A)}_\alpha \, dx^\alpha $
[see Eqs.~(\ref{one_forms1})  and (\ref{one_forms2})].
Following Ref.~\cite{Reb_Tiomno} it is straightforward to show that a scalar field
of the form $\Phi (z)= \varepsilon \,z + \text{const}$ satisfies the scalar field equation
$\Box \,\Phi = \eta^{AB}_{}\,\nabla_{A} \nabla_{B} \,\Phi\,=0$
for a constant amplitude $\varepsilon$ of $\Phi (z)$. Thus, the non-vanishing
components of energy-moment tensor for this scalar field are
\begin{equation}  \label{S-comp}
T^{(S)}_{00} = - T^{(S)}_{11} = - T^{(S)}_{22} = T^{(S)}_{33} = \frac{\varepsilon^2}{2}\,,
\end{equation}
and the field equations (\ref{G_AB}) can be written in the form
\begin{eqnarray}
(3\omega^2 - m^2)f_R + \frac{f}{2} &=& 0 \label{1st-phi}\,, \\
\omega^2f_R - \frac{f}{2} & = & 0 \,,\label{2nd-phi} \\
(m^2 - \omega^2)f_R - \frac{f}{2} &=&  k^2 \varepsilon^2 \,.\label{3rd-phi}
\end{eqnarray}

Equations (\ref{1st-phi}) and (\ref{2nd-phi}) yield
$(4\omega^2- m^2)f_R = 0$, which leads to $m^2=4\omega^2$
for any Palatini $f(R)$ theories that satisfy the WEC condition
$f_R>0$.
This give rise to the unique class of G\"odel-type solutions
\begin{eqnarray}
m^2 &=& 4 \omega^2 \label{1st-phi-a}\,, \\
f_R  & = & \frac{\kappa^2 \varepsilon^2}{2 \,\omega^2 } \,,\label{2nd-phi-b} \\
f &=&  k^2 \varepsilon^2 \,,\label{3rd-phi-c}
\end{eqnarray}
where $f$ and $f_R$ are to be evaluated at $R=2(m^2 - \omega^2)= 3m^2/2$.
{}From equations (\ref{r-critical}) and (\ref{1st-phi-a}) one
clearly has that the critical radius $r_c \rightarrow \infty$.
Hence, for this unique solution there is no violation of causality
of  G\"odel type for any Palatini $f(R)$ gravity with $f_R> 0$.

Finally, we note that the Palatini $f(R)$ theory of form~(\ref{first-f})
permits this unique causal solution for the values of $n$ and $\alpha$
allowed  by the above dynamical systems plus observational data analyses.
Indeed, Eq.~(\ref{first-f}) along with Eqs.~(\ref{1st-phi-a})~--~(\ref{3rd-phi-c})
give
\begin{equation}
m^2 = \frac{2}{3}\,\left[\frac{\alpha}{2}(n+3)\right]^{1/(n+1)},
\end{equation}
which shows that $m^2$ is independent of the amplitude of the scalar field.
Thus,  for $n\in[-0.3,0.3]$ and, for example,  $\alpha = 3.45$ (the best fit value
found in~\cite{Santos}) one has $m \in [2.5,1.6]$. For any other value of $\alpha$
allowed by observations we obviously have a different range of values for $m$
but again with no breakdown of causality of G\"odel-type.

\section{Final Remarks}

A  good deal of effort  has recently gone into the study of
the so-called $f(R)$ gravity. This is motivated by 
the fact that these theories provides an alternative way to
explain the late accelerating expansion of the Universe without invoking either
dark energy matter component or the existence of an extra spatial dimension.
If gravity is governed by a $f(R)$ a number of issues should
be reexamined in the $f(R)$ framework. This includes, for
example, solar system tests, a correct Newtonian limit,
gravitational waves, black holes, four distinct phases in the
evolution history of the Universe, and the breakdown of
causality at a non-local scale.

The underlying space-time manifolds of $f(R)$ gravity theories are assumed
to be locally Lorentzian. Thus, in both formulation of $f(R)$ gravity
the causal structure of the space-time has the same local properties 
of the flat space-time of special relativity, and hence the causality
principle is locally satisfied. The nonlocal question, however, is left
open, and violation of causality can come about.
The general relativity G\"odel model~\cite{Godel49} is the best
known example of a cosmological solution of in which causality
is violated at a nonlocal scale.

In this  paper, we have examined G\"odel-type models and the violation
of causality problem in Palatini $f(R)$ gravity generalizing the results
of Refs.~\cite{RS-2009,CliftonBarrow2005,Reb_Tiomno}.
For physically well-motivated perfect-fluid matter source, we have shown
that every solution with arbitrary $\rho$ that satisfies the weak energy
condition $\rho+p \geq 0$ (or equivalently $df/dR >0$) is
necessarily isometric to the G\"odel geometry, making explicit
that the violation of causality is a generic feature of Palatini $f(R)$
gravity theories.
This extends to the context of Palatini $f(R)$ the
Bampi-Zordan theorem~\cite{BampiZordan78} which was previously
established in the context of Einstein's theory, and has been extended
recently to the framework of $f(R)$ in the metric formalism~\cite{RS-2009}.
We have derived an expression for the critical
radius $r_c$ (beyond which the causality is violated) for an arbitrary
Palatini $f(R)$ theory that satisfies the WEC condition $f_R \geq 0$
making apparent that the violation of causality depends upon both
the $f(R)$ gravity theory and the matter source components ($\rho,p$).

We concretely studied G\"odel-type perfect-fluid solutions in the
specific $f(R) = R - \alpha/R^{n}$ Palatini gravity theory.
We showed that, for positive matter density (with $\rho+p>0$) and
for $\alpha$ and $n$ within the interval permitted by the observational
data,  this theory does not admit G\"odel geometry as a solution of its
field equations.
In this sense, this theory remedies the causal anomaly of G\"odel type
which is allowed in general relativity.
We have also examined the violation of causality of G\"odel type
by considering a scalar field as a matter source. For this source
we showed that any Palatini $f(R)$ gravity gives rise to a unique
G\"odel-type solution with no violation of causality.

\acknowledgments
The authors  acknowledge the support of FAPERJ under a CNE grant.
M.J.R. also thanks CNPq for the grant under which this work
was carried out.
J.S. thanks financial support from Instituto Nacional de Estudos do
Espa\c{c}o (INEspa\c{c}o), Funda\c{c}\~ao de Amparo a Pesquisa do RN (FAPERN),
and the Grant No. 479469/2008-3 from Conselho Nacional de Pesquisa (CNPq).
We are grateful to A.F.F. Teixeira for indicating omissions and misprints.


\begin{thebibliography}{}

\bibitem{theory}
  S. Capozziello, S. Nojiri, S.D. Odintsov and A. Troisi, {Phys. Lett. B} {\bf 639}, 135 (2006);
  S.~Nojiri and S.~D.~Odintsov, {Phys. Lett. B} {\bf 652}, 343 (2007);
  L. Amendola, D. Polarski and S. Tsujikawa, {Int. J. Mod. Phys. D} {\bf 16}, 1555 (2007);
  W. Hu and I. Sawicki, {Phys. Rev. D} {\bf 76}, 064004 (2007);
  C.G. B\"{o}hmer, T. Harko and F.S.N. Lobo, {JCAP} {\bf 0803}, 024 (2008);
  S.A. Appleby and R.A. Battye, {JCAP} {\bf 0805}, 019 (2008);
  T. P. Sotiriou, S. Liberati and V. Faraoni, {Int. J. Mod. Phys. D} {\bf 17}, 399 (2008);
  T.P. Sotiriou, {Phys. Lett. B} {\bf 664}, 225-228 (2008); 
  C.S.J. Pun, Z. Kov\'acs and T. Harko, {Phys. Rev. D} {\bf 78}, 024043 (2008); 
  S.~Nojiri and S.~D.~Odintsov, {Phys. Rev. D} {\bf 78}, 046006 (2008);
  G. Cognola, E. Elizalde, S. Nojiri, S.D. Odintsov, P. Tretyakov and S.Zerbini, {Phys. Rev. D} {\bf 79}, 044001 (2009).
  
\bibitem{Francaviglia}
  S.~Capozziello and M.~Francaviglia, Gen. Rel. Grav. {\bf 40}, 357 (2008);
  T.~P.~Sotiriou and V.~Faraoni, Rev. Mod. Phys. {\bf 82}, 451 (2010); 
  A.~De Felice and S.~Tsujikawa, arXiv:1002.4928 [gr-qc].

\bibitem{Chiba2003} T. Chiba,  Phys. Lett. B {\bf 575}, 1 (2003);
L. Amendola and S. Tsujikawa, Phys. Lett. B {\bf 660}, 125 (2008);
O.M. Lecian and G. Montani, Class. Quantum Grav. {\bf 26}, 045014 (2009).

\bibitem{Sotiriou2006a}
  S.~Nojiri and S.~D.~Odintsov, {Phys. Rev. D} {\bf 68}, 123512 (2003);
  R. Dick, Gen. Rel. Grav. {\bf 36}, 217 (2004);
  T.P. Sotiriou, Phys. Rev. D {\bf 73}, 063515 (2006);
  T.P. Sotiriou, Gen. Rel. Grav. {\bf 38}, 1407 (2006).

\bibitem{Dolgov2003} A.D. Dolgov and M. Kawasaki, Phys. Lett. B {\bf 573},
1 (2003);
V. Faraoni, Phys. Rev. D {\bf 74}, 104017 (2006);
T.P. Sotiriou, Phys. Lett. B {\bf 645}, 389 (2007).

\bibitem{Frolov2008}
  M.~C.~B.~Abdalla, S.~Nojiri and S.~D.~Odintsov, {Class. Quantum Grav.} {\bf 22}, L35 (2005);
  A.V. Frolov, Phys. Rev. Lett. \textbf{101}, 061103 (2008).

\bibitem{energy_conditions}
  J.H. Kung, Phys. Rev. D {\bf 53}, 3017 (1996);
  S.E.P. Bergliaffa, Phys. Lett. B {\bf 642}, 311 (2006);
  J. Santos,  J.S. Alcaniz, M.J. Rebou\c{c}as and F.C. Carvalho, Phys. Rev. D
  {\bf 76}, 083513 (2007); 
  K. Atazadeh, A. Khaleghi, H. R. Sepangi and Y. Tavakoli, Int. J. Mod. Phys. D
  {\bf 18}, 1101 (2009);  
  O. Bertolami and M.C. Sequeira, Phys. Rev. D {\bf 79}, 104010 (2009);  
  J. Santos, M.J. Rebou\c{c}as and J.S. Alcaniz, arXiv:0807.2443[astro-ph].

\bibitem{Amarzguioui}
M. Amarzguioui, \O. Elgar\o y, D.F. Mota and T. Multam\"{a}ki, Astron. Astrophys.
{\bf 454}, 707 (2006).

\bibitem{Tavakol}
  S. Fay, R. Tavakol and S. Tsujikawa, {Phys. Rev. D} {\bf 75}, 063509 (2007).

\bibitem{Santos}
  J. Santos, J. S. Alcaniz, F. C. Carvalho, N. Pires, Phys. Lett. B {\bf 669}, 14 (2008).

\bibitem{Carvalho}
  F.C. Carvalho, E.M. Santos, J.S. Alcaniz and J. Santos, 
  JCAP {\bf 09}, 08 (2008). 

\bibitem{Li}
  B. Li, K. C. Chan and M.-C. Chu, {Phys. Rev. D} {\bf 76}, 024002 (2007).

\bibitem{Yang}
  X.-J. Yang and D.-M. Chen, MNRAS {\bf 394}, 1449 (2009).

\bibitem{Borowiec}
  A. Borowiec, W. God{\l}owski and M. Szyd{\l}owski, {Phys. Rev. D} {\bf 74}, 043502 (2006);
  B. Li and M.-C. Chu, {Phys. Rev. D} {\bf 74}, 104010 (2006);
  T. Koivisto, Phys. Rev. D {\bf 76}, 043527 (2007);
  H. Oyaizu, M. Lima and W. Hu, {Phys. Rev. D} {\bf 78}, 123524 (2008);
  C.-B. Li, Z.-Z. Liu and C.-G. Shao, {Phys. Rev. D} {\bf 79}, 083536 (2009).

\bibitem{Godel49} K. G\"odel,  Rev.\  Mod.\ Phys.\ {\bf 21}, 447 (1949).

\bibitem{BampiZordan78} F. Bampi and C. Zordan, Gen. Rel. Grav. {\bf 9}, 393 (1978).


\bibitem{GT_in_GR}
M.M. Som and A.K. Raychaudhuri,  Proc. Roy. Soc. London A
{\bf 304}, 81 (1968);
A.K. Raychaudhuri and S.N. Guha Thakurta,
 Phys.\ Rev.\  D {\bf 22}, 802 (1980);
M.J. Rebou\c{c}as, J. E. {\AA}man and A.F.F. Teixeira, J. Math. Phys.
{\bf 27}, 1370 (1986);
M.J. Rebou\c{c}as and A.F.F. Teixeira, Phys. Rev. D {\bf 34}, 2985 (1986);
F.M. Paiva, M.J. Rebou\c{c}as and A.F.F. Teixeira, Phys. Lett. A {\bf 126},
168 (1987);
A. Krasi\'nski, J. Math. Phys. {\bf 39}, 2148 (1998);
M. Tsamparlis, D. Nikolopoulos and P.S. Apostolopoulos, Class. Quantum Grav.
{\bf 15}, 2909 (1998);
S. Carneiro, Phys. Rev. D {\bf 61}, 083506 (2000);
Y.N. Obukhov,  arXiv:0008106 [astro-ph];
S. Carneiro and G.A. Mena Marugan, Phys. Rev. D {\bf 64}, 083502 (2001);
J.D. Barrow and C.G. Tsagas, Class. Quantum Grav. {\bf 21}, 1773 (2004);
M.P. Dabrowski and J. Garecki, Phys. Rev. D {\bf 70}, 043511 (2004);
F.C. Sousa, J.B. Fonseca and C. Romero, Class. Quantum Grav. {\bf 25}, 035007 (2008).


\bibitem{GT_Other_Th}
L.L. Smalley,  Phys.\ Rev.\ D {\bf 32}, 3124 (1985);
J. Duarte de Oliveira, A.F.F. Teixeira and J. Tiomno, Phys. Rev. D {\bf 34},
3661 (1986);
A.J. Accioly and G.E.A. Matsas, Phys. Rev. D {\bf 38}, 1083 (1988);
J.D. Barrow and M.P. Dabrowski, Phys. Rev. D {\bf 58}, 103502 (1998);
J.E. {\AA}man, J.B. Fonseca-Neto, M.A.H. MacCallum and M.J. Rebou\c{c}as,
Class. Quantum  Grav. {\bf 15}, 1089 (1998);
M.J. Rebou\c{c}as and A.F.F. Teixeira, J. Math.\ Phys. {\bf 39}, 2180 (1998);
M. J. Rebou\c{c}as and A.F.F. Teixeira, Int.\ J. Mod. Phys. A {\bf 13}, 3181
(1998);
P. Kanti and C.E. Vayonakis, Phys. Rev. D {\bf 60}, 103519 (1999);
H.L. Carrion, M.J. Rebou\c{c}as and A.F.F. Teixeira, J. Math. Phys. {\bf 40},
4011 (1999);
E.K. Boyda, S. Ganguli, P. Horava and U. Varadarajan, Phys. Rev. D {\bf 67},
106003 (2003);
J.D. Barrow and  C.G. Tsagas, Phys. Rev. D {\bf 69}, 064007 (2004);
M. Banados, G. Barnich, G. Compere and A. Gomberoff, Phys. Rev. D {\bf 73},
044006 (2006);
W.-H. Huang, Phys. Lett. B {\bf 615}, 266 (2005); 
D. Astefanesei, R.B. Mann and E. Radu, JHEP {\bf 0501}, 049 (2005);
Y.Brihaye, J. Kunz and E. Radu, JHEP {\bf 08}, 025 (2009); 
C. Furtado, T. Mariz, J.R. Nascimento, A. Yu. Petrov and A.F. Santos,
Phys. Rev. D \textbf{79}, 124039 (2009).


\bibitem{RS-2009}
   M.J. Rebou\c{c}as and J. Santos, Phys. Rev. D {\bf 80}, 063009 (2009).  

\bibitem{CliftonBarrow2005}
   T. Clifton and J. D. Barrow,  Phys. Rev. D {\bf 72}, 123003 (2005).

\bibitem{Reb_Tiomno} M. J. Rebou\c{c}as and J. Tiomno,  Phys.\
Rev.\ D {\bf 28}, 1251 (1983).


\bibitem{TeiRebAman} A.F.F. Teixeira, M.J. Rebou\c{c}as and J.E. Aman,
Phys. Rev. D {\bf 32}, 3309 (1985).

\bibitem{RebAman} M.J. Rebou\c{c}as and J.E. {\AA}man,
 J.\  Math.\ Phys. {\bf 28}, 888 (1987).

\bibitem{Topology} M. Lachi\`{e}ze-Rey and J.P. Luminet, Phys. Rep., \textbf{254}, 135
(1995); J. Levin, Phys.\ Rep.\, \textbf{365}, 251 (2002);
M.J. Rebou\c{c}as and G.I. Gomero, Braz. J. Phys. \textbf{34}, 1358 (2004);
G.I. Gomero, M.J. Rebou\c{c}as, and  R. Tavakol, Class. Quantum
Grav. \textbf{18}, 4461 (2001); G.I. Gomero, M.J. Rebou\c{c}as, and
R. Tavakol, Class. Quantum Grav. \textbf{18}, L145 (2001).

\bibitem{EC}
S.W. Hawking and G.F.R. Ellis, {\em The Large Scale Structure of Spacetime},
(Cambridge University Press, England, 1973);
S. Carroll, {\em Spacetime and Geometry: An Introduction to General Relativity},
(Addison Wesley, New York, 2004);
R.M. Wald, \emph{General Relativity}, (University of Chicago Press, Chicago, 1984).

\bibitem{EC-observation}
M. Visser, \emph{Science} \textbf{276}, 88 (1997);
M. Visser, Phys. Rev. D \textbf{56},  7578 (1997);
J. Santos, J.S. Alcaniz and M.J. Rebou\c{c}as, Phys. Rev. D \textbf{74}, 067301 (2006);
J. Santos, J.S. Alcaniz, N. Pires and M.J. Rebou\c{c}as, Phys. Rev. D \textbf{75}, 083523 (2007);
J. Santos, J.S. Alcaniz, M.J. Rebou\c{c}as and N. Pires, Phys. Rev. D \textbf{76}, 043519 (2007);
M.P. Lima, S. Vitenti and M.J. Rebou\c{c}as, Phys. Rev. D \textbf{77}, 083518 (2008);
M.P. Lima, S.D.P. Vitenti and M.J. Rebou\c{c}as, Phys. Lett. B \textbf{668}, 83 (2008);
M.P. Lima, S.D.P. Vitenti and M.J. Rebou\c{c}as, arXiv:0812.0811v2 [astro-ph].

\bibitem{Amendola}
L. Amendola, D. Polarski and S. Tsujikawa, Phys. Rev. Lett. {\bf 98}, 131302 (2007).




\end{thebibliography}
\end{document}